# Photoacoustic Imaging using Combination of Eigenspace-Based Minimum Variance and Delay-Multiply-and-Sum Beamformers

## Simulation Study

Moein Mozaffarzadeh, Seyed Amin Ollah Izadi Avanji
Department of Computer and Electrical Engineering,
Biomedical Engineering
Tarbiat Modares University
Tehran, Iran
moein.mfh@modares.ac.ir, amin.izadi@modares.ac.ir

Ali Mahloojifar*, Mahdi Orooji
Department of Computer and Electrical Engineering,
Biomedical Engineering
Tarbiat Modares University
Tehran, Iran
mahlooji@modares.ac.ir, morooji@modares.ac.ir

*Abstract*—**Delay and Sum (DAS), as the most common beamforming algorithm in Photoacoustic Imaging (PAI), having a simple implementation, results in a low-quality image. Delay Multiply and Sum (DMAS) was introduced to improve the quality of the reconstructed images using DAS. However, the resolution improvement is now well enough compared to high resolution adaptive reconstruction methods such as Eigenspace-Based Minimum Variance (EIBMV). We proposed to integrate the EIBMV inside the DMAS formula by replacing the existing DAS algebra inside the expansion of DMAS, called EIBMV-DMAS. It is shown that EIBMV-DMAS outperforms DMAS in the terms of levels of sidelobes and width of mainlobe significantly. For instance, at the depth of 35 mm, EIBMV-DMAS outperforms DMAS and EIBMV in the term of sidelobes for about 108 dB, 98 dB and 44 dB compared to DAS, DMAS, and EIBMV, respectively. The quantitative comparison has been conducted using Full-Width-Half-Maximum (FWHM) and Signal-to-Noise Ratio (SNR), and it was shown that EIBMV-DMAS reduces the FWHM about 1.65 mm and improves the SNR about 15 dB, compared to DMAS.**

## I. INTRODUCTION

Photoacoustic imaging (PAI) is an emerging medical imaging modality providing the resolution of Ultrasound (US) imaging and the contrast of optical imaging [1]. In this imaging modality, US signals are generated based on the thermoacoustic effect as a result of laser illumination and US transducers are used to detect the propagated signals [2]. PAI uses the optical absorption distribution to provide functional, structural and anatomical information, and it can be used in different applications such as tumor detection [3], ocular imaging [4], monitoring oxygenation in blood vessels and functional imaging [5]. There are two method is imaging in PAI: Photoacoustic Tomography (PAT) and Photoacoustic Microscopy (PAM) [6, 7]. In PAT, US array of transducers in different shapes such as linear, circular and arc, are used in order to detect the generated Photoacoustic (PA) signals, and a reconstruction algorithm is used to form the optical absorption distribution map of the tissue being imaging [8, 9]. There are some simplification assumptions in these reconstruction algorithms resulting in artifacts in the reconstructed PA images. Nowadays, removing these negative effects is one of the most important challenges in PAI [10]. As one of the PAI techniques, PAM has been significantly investigated in field of biology. Beyond the material of tissue, the axial and lateral resolution of PAM depends on numerical aperture and bandwidth of US transducer used for PA wave's detection called Acoustic-Resolution PAM (AR-PAM) [11]. There are a high similarity between the PA and US signals, and the reconstruction algorithms can be used for both of these imaging modalities with some modifications [12]. There are a number of researches focused on the use of a single image formation method for an integrated PA-US imaging system. Consequently, the cost of imaging system will be reduced [13].

The most commonly used beamforming algorithm for linear array US and PA imaging is Delay and Sum (DAS) due to its simple implementation and real-time imaging. However, it results in a low-resolution image along with high levels of sidelobes. Matrone et al. proposed, in [14], a new beamforming algorithm namely Delay-Multiply-and-Sum (DMAS) which was used as a reconstruction algorithm in confocal microwave imaging for breast cancer detection [15]. This algorithm has been modified and was reported by recent publications of our research group [16, 17]. In this paper, it is proposed to combine Eigenspace-Based Minimum Variance (EIBMV) adaptive beamforming algorithm, [18], with the DMAS method. It is shown that although DMAS improves the quality of the reconstructed PA image, it results in a low-resolution image. Using the proposed method, achieving a high resolution image along with a lower levels of sidelobes, compared to DMAS and EIBMV, would be available.

---

*Corresponding Author

The rest of the paper is as follows. Section II contains the theory of the beamformers. The proposed method is introduced in section III. Numerical simulation along with results and performance assessment are presented in section IV. Finally, the conclusion is presented in section VII.

## II. BACKGROUND

### 1) Photoacoustic Theory

In typical PAI, waves are propagated based on the thermoelastic expansion and US transducers detect the PA signals. Under thermal confinement, acoustic homogeneous medium and inhomogeneous optical absorption medium the pressure $P(r,t)$ at position r and time t, results from heat sources $H(r,t)$ and obeys the following equation:

$$c^2 \nabla^2 P(r,t) - \frac{\partial^2}{\partial t} P(r,t) = -\frac{\Gamma(r) H(r,t)}{t}, \quad (1)$$

Where $\Gamma(r) = \beta c^2 / C_p$ is the gruneisen parameter, $\beta$ is the isobaric volume expansion, $c$ is the speed of sound and $C_p$ is the heat capacity [7]. The heat function can be written as the product of two components:

$$H(r,t) = A(r) I(t), \quad (2)$$

Where $A(r)$ is the spatial absorption function, and $I(t)$ is the temporal illumination function [7]. Assuming $I(t) = \delta(t)$, the detected acoustic pressure $P(r_0, t)$ at the detector position $r_0$ and the time $t$ can be written as:

$$P(r_0, t) = \frac{1}{c} \frac{\partial}{\partial t} \iiint d^3 r \, D(r) \frac{\delta(ct - |r_0 - r|)}{4\pi |r_0 - r|}, \quad (3)$$

Where $D(r) = \Gamma(r) A(r)$. (3) is called forward problem in PAI. The reconstruction of $A(r)$ from the detected PA waves is considered as the backward problem. Based on the condition explained in [19], the PA image can be reconstructed as follows:

$$D(\rho, \phi, z) = -\frac{1}{2\pi c^2} \iint ds_0 [n. n_0] \frac{1}{t} \frac{\partial P(r_0, t)}{\partial t} \Big|_{t = \frac{|r - r_0|}{c}}, \quad (4)$$

Where

$$n. n_0 = \frac{(\rho - \rho_0)}{(r - r_0)} = \sqrt{\frac{\rho^2 + \rho_0^2 - 2\rho\rho_0 \cos(\phi - \phi_0)}{(r - r_0)^2}}$$

$$= \sqrt{1 - \frac{(z_0 - z)^2}{(r - r_0)^2}}, \quad (5)$$

$ds_0 = \rho_0 d\phi_0 dz_0$ and $\rho_0$ are the projection of $r$ and $r_0$ on the $z$ plane [19]. In PA image reconstruction, (4) is called back-projection (PB).

### 2) Beamforming

As the reconstruction algorithm, DAS can be used to form the PA image using the detected signals:

$$y_{DAS}(k) = \sum_{i=1}^{M} x_i(k - \Delta_i), \quad (6)$$

Where $y_{DAS}$ is the output of beamformer, M is number of array elements, $x_i(k)$ and $\Delta_i$ are detected signals and corresponding time delay for element $i$, respectively [20]. (6) results in a low-quality image. DMAS was introduced by Matrone et al. in [14], and it was shown that it can be used to improve the quality of the formed images compared to DAS. DMAS formula is as follows:

$$y_{DMAS}(k) = \sum_{i=1}^{M-1} \sum_{j=i+1}^{M} x_j(k - \Delta_j) x_j(k - \Delta_j). \quad (7)$$

To overcome the dimensionally squared problem of (7) following equations are suggested [14]:

$$\hat{x}_{ij}(k) = sign[x_j(k - \Delta_j) x_j(k - \Delta_j)].$$
$$\sqrt{|x_j(k - \Delta_j) x_j(k - \Delta_j)|}, \quad for \ 1 \leq i \leq j \leq M, \quad (8)$$

$$y_{DMAS}(k) = \sum_{i=1}^{M-1} \sum_{j=i+1}^{M} \hat{x}_{ij}(k). \quad (9)$$

More information and explanation about the DMAS algorithm can be seen in [14], and it is out of the scope of this paper. Although it was proved that DMAS outperforms DAS in the terms of resolution and levels of sidelobes, its resolution is not satisfying in comparison with Minimum Variance-Based algorithms [18, 20]. In the following, it is shown that we can integrated the EIBMV adaptive beamformer inside the DMAs formula expansion.

## III. PTOPOSED METHOD

In this paper, the goal is to integrate the EIBMV beamformer inside the DMAS algorithm to improve the image quality. Expanding (7) results in several terms as follows:

$$y_{DMAS}(k) =$$
$$[x_{d1}(k) x_{d2}(k) + x_{d1}(k) x_{d3}(k) \dots + x_{d1}(k) x_{dM}(k)]$$
$$+ \quad [x_{d2}(k) x_{d3}(k) + x_{d2}(k) x_{d4}(k) \dots + x_{d2}(k) x_{dM}(k)]$$
$$+ \quad \dots$$
$$+ \quad [x_{d(M-1)}(k) x_{dM}(k)]. \quad (11)$$

Using factorization, the following formula can be obtained:

$$y_{DMAS}(k) =$$
$$x_{d1}(k)[x_{d2}(k) + x_{d3}(k) \dots + x_{dM}(k)]$$
$$+ \quad x_{d2}(k) [x_{d3}(k) + x_{d4}(k) \dots + x_{dM}(k)]$$

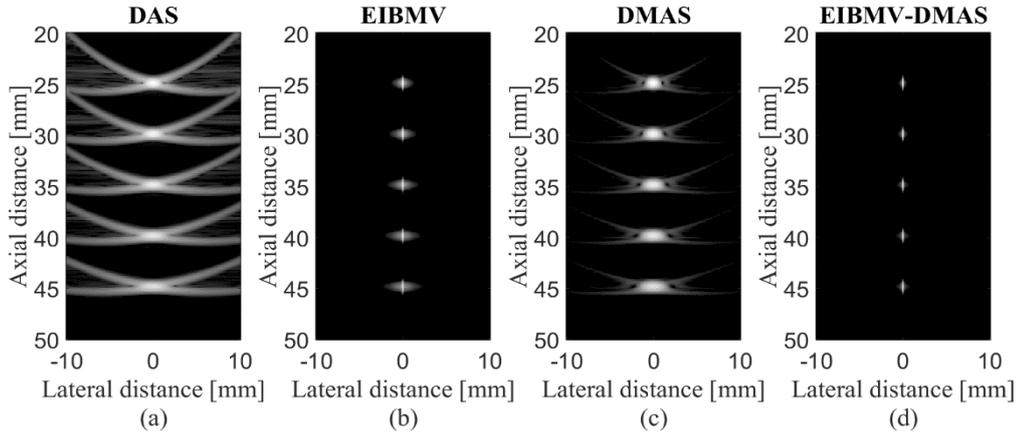

Figure 1. The reconstructed images using (a) DAS, (b) EIBMV, (c) DMAS, and (d) EIBMV-DMAS. All the images are shown with a dynamic range of 60 dB. The noise was added to the detected signals considering a SNR of 50 dB.

$$+ \quad \dots$$
$$+ \quad x_{d(M-1)}(k) \, [x_{dM}(k)]. \qquad (12)$$

As can be seen in (12), which is mathematically equal to (11), there are summation in each bracket interpreting as a DAS algorithm. It was proved that DAS leads to a low-resolution image and high sidelobes. Consequently, the existing of DAS inside the DMAS algebra expansion can be the reason of the low resolution of DMAS compared to MV-Based algorithms. It is proposed to use EIBMV algorithm instead of the outer summation procedure in the expansion of DMAS algorithm, (12), to improve the resolution of the reconstructed images, and we call it EIBMV-DMAS in this paper. It should be noticed that since there is multiplication and consequent square root in the DMAS algorithm, we face them in the proposed method too. In [14], a band-pass filter was used, considering the central frequency of the imaging system, in order to pass the necessary information. In the proposed method, the necessary band-pass filter is implemented on the each part of the outer summation of (12), and then the EIBMV is applied. Moreover, the (8) and (9) are used to solve the square root. In the following section, it can be seen that EIBMV-DMAS outperforms DAS, DMAS and EIBMV in the terms of the resolution and levels of sidelobes.

## IV. Results

K-wave Matlab toolbox was used to design numerical study [21]. Five 0.1 mm radius spherical absorbers as initial pressure were positioned along the vertical axis every 5 mm, beginning distance of 25 mm from transducer surface. Imaging region was 20 mm in lateral axis and 50 mm in vertical axis. A linear array with M=128 elements operating at 4 MHz center frequency and 77% fractional bandwidth, was used to detected the PA signals. Speed of sound was assumed to be 1540 m/s, and sampling frequency was 50MHz. Gaussian noise was added to detected signals where SNR of the signals is 50 dB. The subarray length $L = M/2$, $K = 5$, $\sigma = 0.5$, and $\Delta = 1/10L$ for all the simulations. Also, a band-pass filter was applied by a Tukey window ($\alpha = 0.5$) to the beamformed signal spectra, covering 4-12 MHz, to pass the necessary information. The reconstructed images using DAS, DMAS, EIBMV and

EIBMV-DMAS, and the related lateral variation at the two depths of imaging are shown in Figure 1 and Figure 2, respectively. As can be seen in Figure 1, the reconstructed image using DAS have a low quality, as expected, and the point targets are not well-detectable.

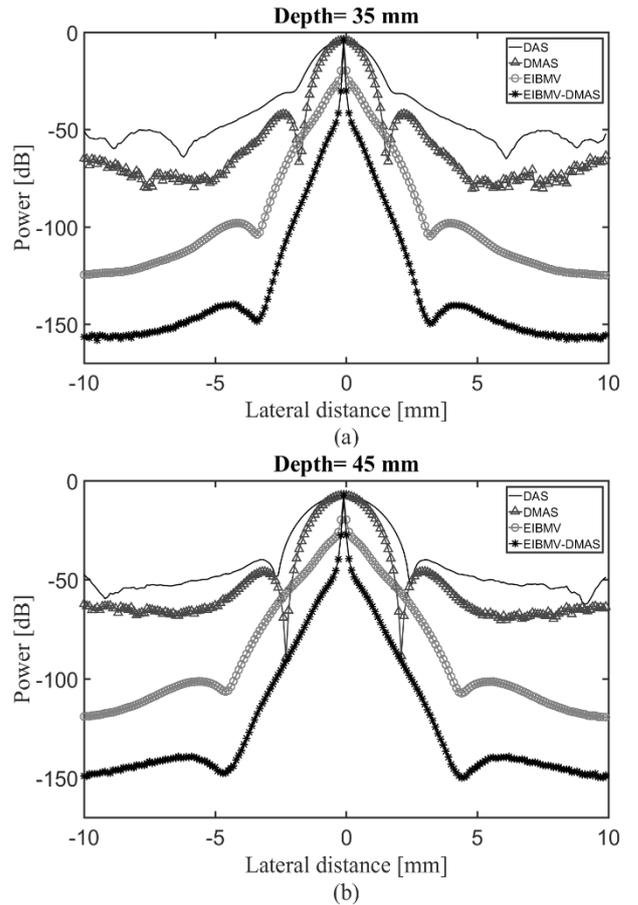

Figure 2. Lateral variations of DAS, DMAS, EIBMV, and EIBMV-DMAS at the depth of (a) 35 mm and (b) 45 mm.

DMAS reduces the levels of sidelobes and artifacts, but the resolution is not significantly improved in comparison with DAS. The EIBMV-DMAS results in lower levels of sidelobes and artifacts compared to DMAS, while the resolution improvement is significant. Moreover, EIBMV-DMAS reduces the sidelobes in comparison with EIBMV while the resolution is retained. Considering the Figure 2, it can be seen that EIBMV-DMAS leads to lower levels of sidelobes and width of mainlobe compared to other beamformers. Consider, for instance, the depth of 35 mm where the sidelobe levels for DAS, DMAS, EIBMV, and EIBMV-DMAS is for about -31 dB, -42 dB, -98 dB, and -140 dB, respectively.

To compare the beamformers in details, quantitative comparison has been conducted using Full-Width-Half-Maximum (FWHM) metric in -6 dB and Signal-to-Noise Ratio (SNR). The calculated FWHM for each beamformer at the depth of 45 mm is presented in TABLE I. As can be seen, the EIBMV-DMAS results in narrower width of mainlobe compared to other beamformers while the improvement is significant in comparison with DAS and DMAS. The calculated SNR, using the method illustrated in reference [17], for each beamformer at the depth of 45 mm is shown in TABLE II. As shown, EIBMV-DMAS outperform in the term of SNR compared to other beamformers, having a higher SNR.

## I. CONCLUSION

In this paper, it was shown that DMAS beamformer results in a low resolution image in comparison with high resolution algorithms such as EIBMV. The expansion of DMAS was used to integrate the EIBMV adaptive beamforming method with the DMAS in order to improve its resolution. The numerical results showed that the  EIBMV-DMAS leads to higher image quality, in comparison with DMAS and EIBMV, having a narrower width of mainlobe and lower levels of sidelobes. Quantitative results indicated that EIBMV-DMAS improves the DMAS in the terms of FWHM and SNR for about 94% and 50%, respectively.

Table 1. FWHM in -6 dB values at the depth of 45 mm for all the beamformers.

| Beamformer | FWHM in -6 dB |
|---|---|
| DAS | 2.41 mm |
| DMAS | 1.75 mm |
| EIBMV | 0.14 mm |
| EIBMV-DMAS | 0.10 mm |

Table 2. SNR (dB) values at the depth of 45 mm for all the beamformers.

| Beamformer | SNR (dB) |
|---|---|
| DAS | 27.37 |
| DMAS | 30.98 |
| EIBMV | 41.61 |
| EIBMV-DMAS | 45.62 |